\documentclass[11pt,a4paper]{article}
\usepackage{jcappub}
\usepackage[sort&compress]{natbib}
\usepackage{hyperref,ifthen}
\begin{document}

\title{The   Dark Matter Density in the Solar Neighborhood reconsidered }

\author{W.~ de Boer,}
\author{M.~ Weber}
\affiliation{Institut f\"ur Experimentelle Kernphysik, Karlsruher Institut
           f\"ur Technologie (KIT), P.O. Box 6980, 76128 Karlsruhe, Germany}

\abstract{
The peculiar dip in the outer rotation curve at a distance of 9 kpc, which was recently confirmed by precise measurements with the VERA VLBI array in Japan, suggests donut-like substructures in the dark matter (DM) halo, since
 spherical or elliptical distributions will not cause a dip.
Additionally, such a donut-like DM structure seems to be required by the dip in the gas flaring of the disk.
In this paper we consider the impact of such DM substructure in the disk on the rotation curve, the gas flaring, the local DM density and the local surface density.
A global fit  shows that the rotation curve is  best described by an NFW DM profile complemented by two donut-like DM substructures at radii of 4.2 and 12.4 kpc, which  coincide with the local dust ring and the Monocerus ring of stars, respectively. Both  regions have been suggested  as regions with tidal streams from ''shredded'' satellites, thus enhancing the plausibility for additional DM.  If real, the radial extensions of these nearby ringlike structures enhance the local dark matter density by a factor of  four to about 1.3$\pm0.3$ GeV/cm$^3$.
We find that i) this  higher DM density is perfectly consistent with the local gravitational potential determining the surface density and ii) the s-shaped gas flaring is explained.  Such a possible  enhancement of the local DM density is of great interest for direct DM searches and the ringlike structure would change the directional dependence of gamma rays for indirect DM searches.
}

\maketitle
\section{Introduction}
The flat rotation curves (RC) of galaxies show that dark matter (DM) is present in the form of large scale halos with a density slowly falling off from the center (like 1/$r^2$ for a flat RC), see e.g. \citep{Fich:1991ej,Sofue:2000jx}.
In our Galaxy the rotation curve is not flat but has two peculiar dips, one at 3 kpc and one at 9 kpc \citep{Sofue:2008wt}.
These dips can not be explained by smoothly varying components, but need ringlike perturbations in the disk, as studied in detail in Ref. \citep{Sofue:2008wt}. These authors combined all available rotation curve data and assumed substructure in the visible sector to explain the dips. However, especially for the ''outer'' ring located outside the solar radius of 8.3 kpc, the amount of visible matter rapidly decreases and it seems hard to explain the increase in RC with a sudden increase in visible matter.

To determine the local DM density one has usually ignored this increase in the RC, assuming that the distance measurements are too uncertain, see e.g. \citep{Binney:1996fb}. The results for the local DM density are then around 0.3-0.4 GeV/cm$^3$ \citep{Catena:2009mf,Weber:2009pt,Salucci:2010qr} with the precise value depending on the assumptions and fitting procedures. E.g. in Ref. \citep{Weber:2009pt} it is shown that the various disk parameters are almost 100\% correlated with the DM profile parameters, which might be at the origin of the rather large differences in error estimates.
A local DM density of 0.3  GeV/cm$^3$ is only 8\% of  the local matter density, as determined from the precise measurements of the gravitational potential using the star counts and velocity measurements from the Hipparcos satellite \citep{Korchagin:2003yk,Holmberg:2004fj,Bienayme:2005py} in agreement with earlier measurements of the surface density \citep{Kuijken:1989,Kuijken:1989hu,Kuijken:1991mw}.
Most of these authors suggested from the decrease in the inner RC combined with the low surface density that there is no need for a significant contribution of DM in our solar neighborhood.
However, there are four additional new pieces of evidence, that bear on this question:
\begin{itemize}
\item
Measurements with the VERA VLBI array in Japan using the maser light from molecular clouds confirm the peculiar change of slope around 9 kpc \citep{Oh:2010zz}.
With the precise distance measurements one cannot ignore the high rotation speed in the outer RC anymore.
\item
The  thickness of the gas layer increases with increasing distance from the Galactic center, the so-called gas flaring. The thickness is given by the gravitational potential and shows a clear ''dip'' in the outer Galaxy, which seems only explainable by
a donut-like DM substructure in the Galactic plane \citep{Kalberla:2007sr}.
\item
The Monocerus ring of stars in the outer Galaxy  was originally discovered in the SDSS data \citep{Newberg:2001sx,Yanny:2003zu}. Subsequently it was found to span  an almost circular structure of low-metallicity stars of at least 170 degrees around the Galaxy in the plane of the disk, see e.g. \citep{Ibata:2003di,RochaPinto:2004ru,Martin:2005xa,Conn:2005xd,Martin:2006rg}.
Such a large overdensity over a large range is suggestive of the infall of a dwarf galaxy and N-body simulations  using the Monocerus ring as input predict DM ringlike structures in the Galactic plane
with the closest and most dense ring located at 13 kpc from the Galactic center, see Refs. \citep{Martin:2003vk,Penarrubia:2004qw,Kazantzidis:2007hy} and references therein.
This leads to DM densities in the plane of the disk in broad agreement with the gas flaring, which forms a strong argument against  interpretations of the Monoceros ring as extensions of the disk, which led to lively discussions, see e.g. \citep{Conn:2005cn,Moitinho:2006ru,Momany:2006ch}.
\item
additional dark matter in the inner region has been suggested from
the early infall of a dwarf galaxy, presumably some 13 billion years ago \citep{Wyse:2006xz,Ruchti:2010cz,Robin:1995fr,Dierickx:2010jc,Kazantzidis:2009zq,Read:2009iv}, which created the thick disk. The thick
disk has a significantly larger scale height and its mainly metal-poor stars imply a different population of old stars.
A possible hint of a ringlike
gravitational potential well in the inner Galaxy is provided by the ringlike structure of the dust ring with the highest density of the dust ring located at a radius of about 4 kpc from the Galactic center and a radial extension of about 1 kpc \citep{Drimmel:2001ti}.
\end{itemize}

It is the purpose of this paper to reinvestigate the local DM density by assuming two donut-like overdensities of matter in the Galactic plane, one inner ring and one outer ring in order to explain the dips in the rotation curve at 3 and 9 kpc as well as the dip  in the gas flaring.
The data on the RC from Ref. \citep{Sofue:2008wt} was supplemented with the
new precise measurements with the VERA VLBI array in Japan using the maser light from molecular clouds \citep{Oh:2010zz}.
The fitting procedure is described in our previous paper, where the new VERA data were not available yet and the dip in the rotation curve at 9 kpc was ignored \citep{Weber:2009pt}; this paper will be referred to as Paper I.
\section{Analysis}
\begin{figure}
  \begin{center}
  \includegraphics[width=0.8\columnwidth,angle=0]{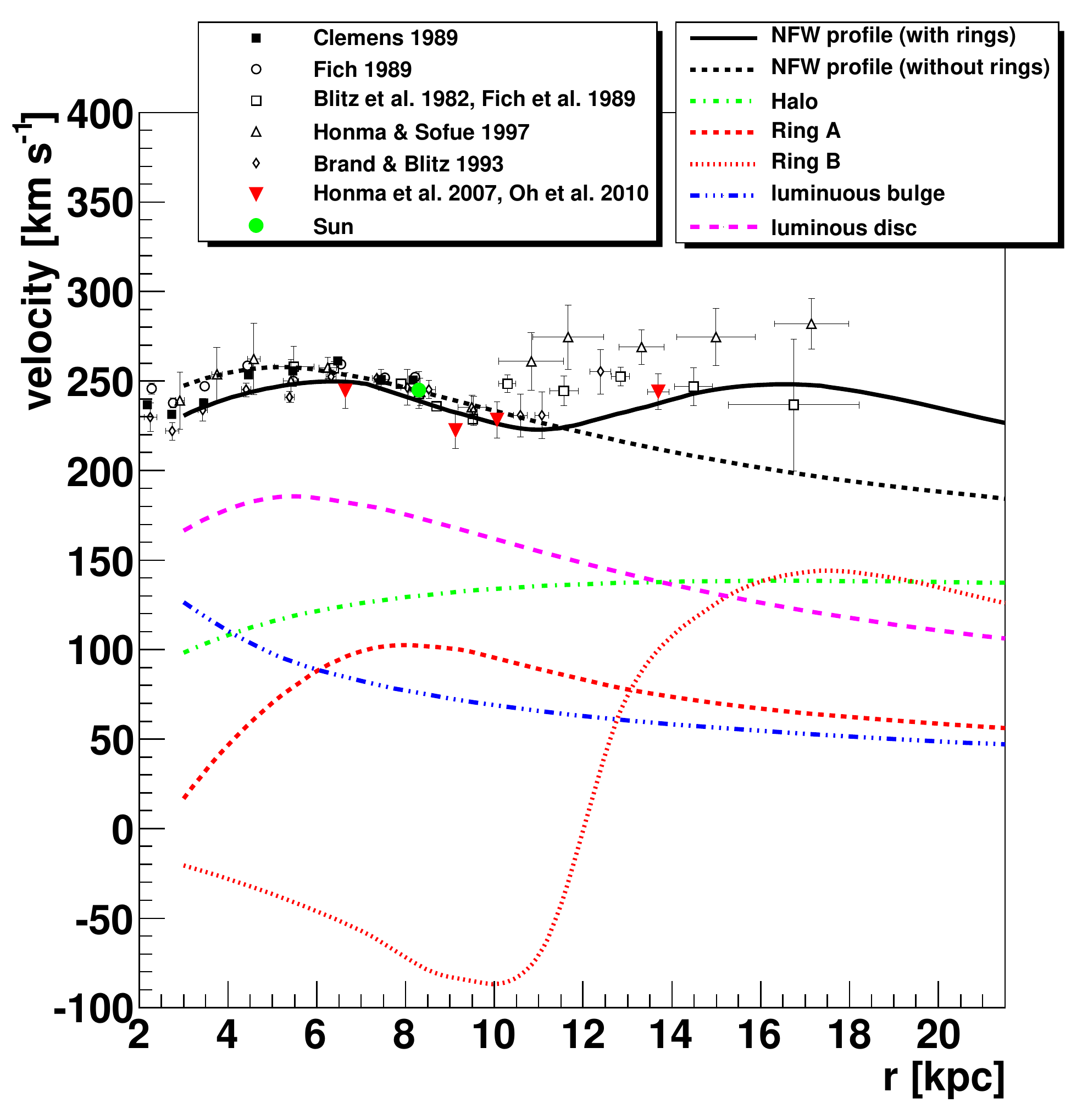}
   \caption[]{The Galactic rotation curve (RC). The dashed line corresponds to a fit of the inner RC
   only with a smooth NFW profile. The solid line includes DM ring-like substructure at 4.2 and 12.4 kpc.
   For clarity the weighted average of  many data points has been plotted, where the data was  taken from Ref. \citep{Weber:2009pt} with in addition the precisely measured ones with VBLI technology from Ref. \citep{Oh:2010zz}, shown as upside-down (red) triangles.}
    \label{f1}
  \end{center}
\end{figure}
\subsection{Rotation curve}
The DM profile  including the rings can  be parameterized as follows:
\begin{eqnarray}
  \nonumber
  \rho_\chi (\vec r) &=& \rho_{\odot,\mathrm{Halo}} \cdot \left ( \frac{\tilde r}{r_\odot} \right )^{-\gamma}
  \cdot \left \lbrack \frac{1 + \left ( \frac{\tilde r}{a} \right )^\alpha}{1 + \left
      ( \frac{r_\odot}{a} \right )^\alpha} \right \rbrack^{\frac{\gamma -
    \beta}{\alpha}}
    \\[1mm]&&
     +\ \sum_{n=1}^2 \rho_n\ \exp \left ( - \frac{(\tilde r_{gc,n}
    - R_n)^2}{2 \cdot \sigma_{R,n}^2} - \left \vert \frac{z}{\sigma_{z,n}}
  \right \vert \right ),
  \label{eq:HaloRings}
\end{eqnarray}
\begin{equation}\nonumber
\tilde r = \sqrt{x^2 + \frac{y^2}{\epsilon_{xy}^2} + \frac{z^2}{\epsilon_z^2}},
\hspace{1cm} \tilde r_{gc,n} = \sqrt{x_{(n)}^2 + \frac{y_{(n)}^2}{\epsilon_{xy,n}}},
\label{rtilde}
\end{equation}
where the first term in Eq. \ref{eq:HaloRings} represents the smooth Galactic halo  and the second term describes the DM donut-like rings. For the smooth DM halo an NFW profile with $\alpha=1, ~\beta=3,~\gamma=1$  in Eq. \ref{eq:HaloRings} and a scale radius of 12 kpc was taken, which yields an excellent fit to the inner rotation curve \citep{Weber:2009pt}, as displayed by the short-dashed line in Fig. \ref{f1}.
In the rings the DM distribution in
z-direction is taken to decrease exponentially with the
scale height $\sigma_{z,n}$ and distributed in r-direction
around the ring radius $R_n$ with a  Gaussian width $\sigma_{R,n}$.
The rings have been assumed to be circular, i.e. the ellipticities $\epsilon_{xy}$ and $\epsilon_{z}$ have been set to 1,
although in reality infall of dwarf galaxies will never lead to perfectly circular rings. However,   the used data is
 most sensitive to the nearest segment of the stream, which seems locally well described by
a circular structure and changing the ellipticity did not improve the fit.
The Gaussian radial width of the outer ring is taken to be slightly different for
the inner and the outer side\footnote{The radial density was taken to decrease at the inner side  as an s-shape built from two parabolic functions, which reach zero
at 5 kpc from the center of the ring, i.e. at R=7.4 kpc. }, which is required for the steep increase of the slope in the rotation curve
after 9 kpc combined with the slow decrease at larger radii.
Such a difference is expected from the conservation
of angular momentum for tidal streams, since they cannot
reach arbitrary small Galactocentric distances.
\begin{table}[]
\begin{center}
\begin{tabular}{|c|c|c|c|c|}
  \hline
  Parameter & Inner Ring & Outer Ring \\\hline
  $\rho_n$ [GeV cm$^{-3}$] & 3.5 & 2.1 \\
  r [kpc]                     & 4.2 & 12.4 \\
  $\sigma_r$ [kpc]   & 2.5 & 3.2 \\
  $\sigma_z$ [pc]    & 450 & 650 \\
\hline
\end{tabular}
\end{center}
\caption[Ring parameters]{Ring parameters of the best $\chi^2$ fit. The errors are typically 30\%, but are correlated with the disk parameters, which have been fixed to the values given in Ref. \citep{Weber:2009pt}. \label{table1} }
\end{table}
\begin{figure}
  \begin{center}
  \includegraphics[width=0.8\columnwidth,angle=0]{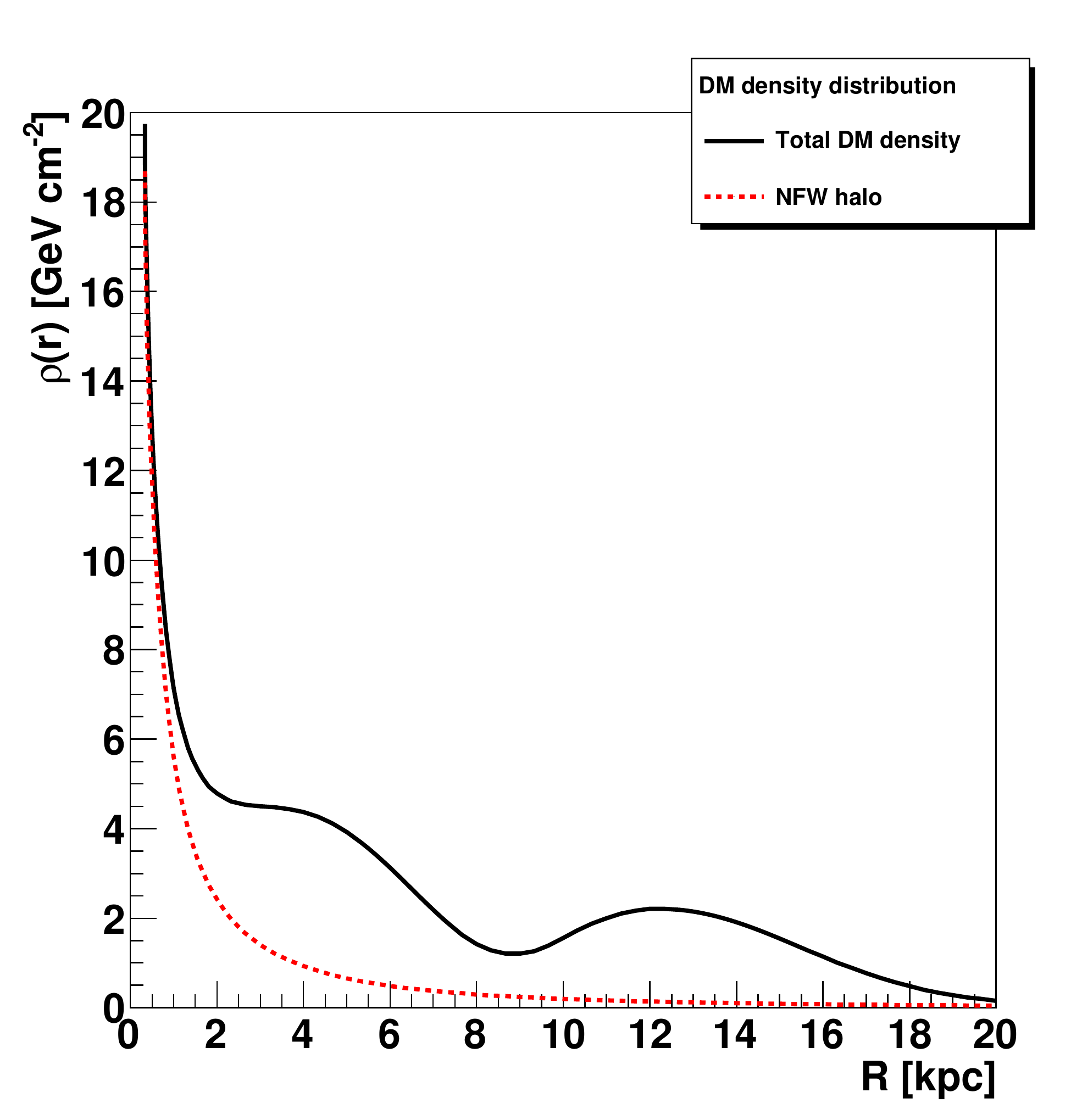}
   \caption{The radial dependence of the DM density including the ringlike structures at 4.2 and 12.4 kpc.}
    \label{f2}
  \end{center}
\end{figure}
The outer RC is much better described,
when the rings are included,  as shown by the solid line in Fig. \ref{f1}. The parameters of the outer ring are  dominated by the precise VLBI data points from VERA. The main reason for the decreasing inner RC in this scenario is {\it not} the lack of DM, but the large negative contribution from the outer ring, which creates an outward gravitational force, thus reducing the centrifugal force and hence the rotation speed inside the ring.
The ring parameters for the best fit to the data
are given in Table \ref{table1}.

These parameters provide DM rings with masses of $1.4 \cdot 10^{10}$ solar masses
for the inner ring and $3.9 \cdot 10^{10}$ solar masses for the outer ring.
Although these masses are only a minor contribution to the total mass of about $10^{12}$ solar masses, one may worry about the disk stability
and probability of accreting such heavy masses.
A recent paper on the assembly history  of  DM halos \citep{Wang:2010ik}
finds that in three of their six simulated galaxies  the central
mass was brought in through mergers involving host halos with mass greater than
$10^{10} h^{-1}$ solar masses in agreement with previous studies \citep{Kazantzidis:2007hy}.
 Furthermore they
state that ''late-accreted mass typically settles in the outskirts of a halo,
thus allowing the hierarchical growth of halos to be reconciled with the
ubiquitous presence of thin stellar disks'', again in agreement with Ref. \citep{Kazantzidis:2007hy}.
Also the thin disk may be regrown from the cooling of the heated gas after a merger, see e.g.   \citep{Villalobos:2009qc}
and if the disk survival depends sensitively on the way how the accreted mass approaches the Galaxy and how it is introduced in the simulations \citep{Kazantzidis:2009zq}.
It should be remembered that our neighboring galaxy, Andromeda (M31), shows also a ringlike structure in the outer galaxy, which is attributed to the infall of a dwarf galaxy as well \citep{Ibata:2005bd}.

The local  density of the smooth NFW halo $\rho_{\odot, \mathrm{Halo}}$ is found to
be 0.28 GeV/cm$^{3}$, while the inner and outer rings contribute
   0.87 and  0.14 GeV~cm$^{-3}$, respectively, which leads to:
\begin{equation}\label{oort1}
    \rho_{\odot,\mathrm{DM}} = 1.3\pm 0.3 ~\mathrm{GeV/cm}^{3} = 0.030\pm 0.007 ~\mathrm{M}_\odot/\mathrm{pc^3},
\end{equation}
 so the rings enhance the local DM density by about a factor four.
The error is systematic and depends mainly on the assumed parameterizations of the rings.
The radial dependence of the  DM density within the Galactic disk is shown in Fig.
\ref{f2}. The influence of the DM rings is clearly
visible.
\begin{figure}
  \begin{center}
  \includegraphics[width=0.8\columnwidth,angle=0]{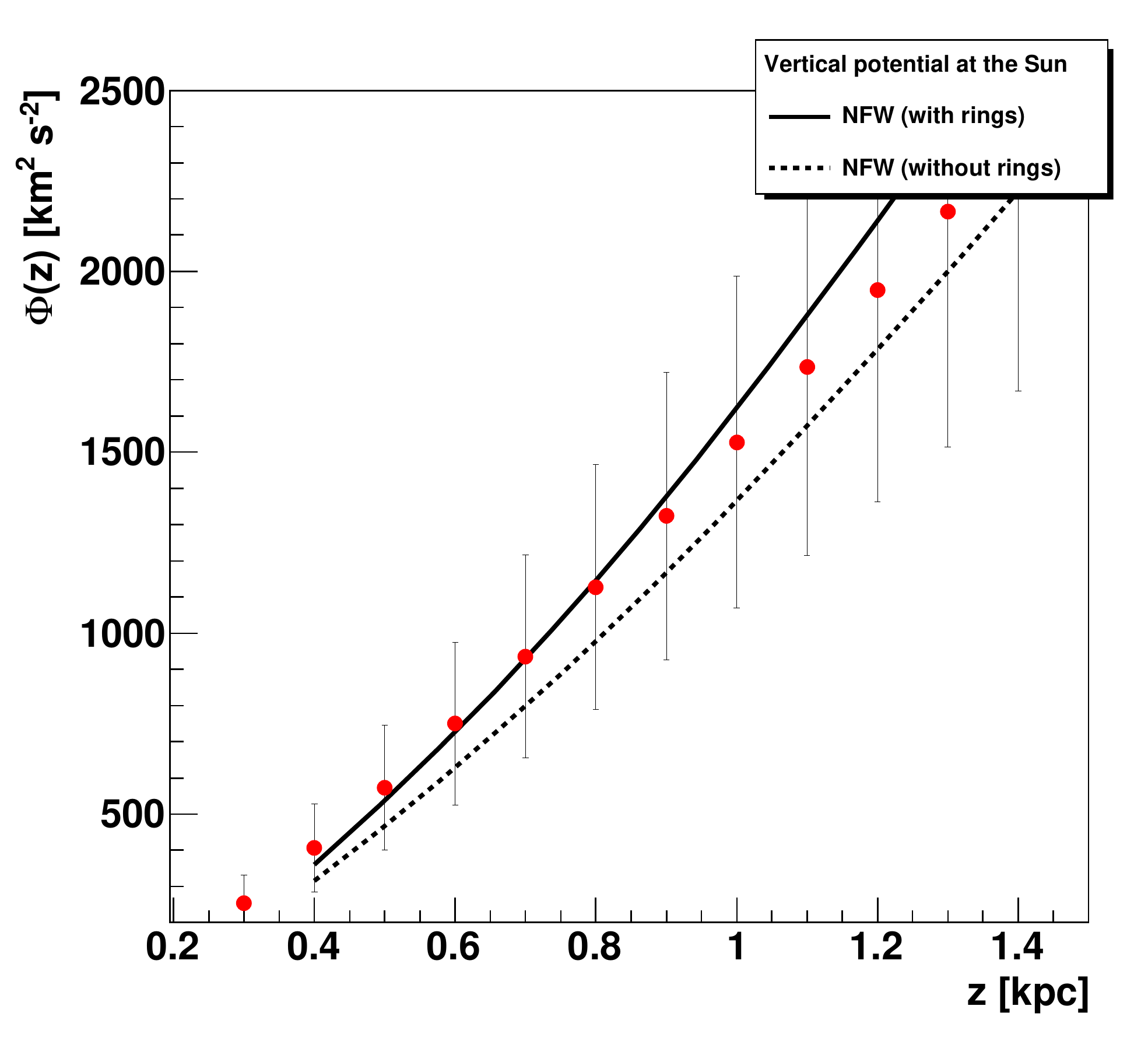}
   \caption[]{The vertical gravitational potential
             at the position of the Sun $\Phi(\mathrm{z})$ for a fit with and without DM ringlike substructures.
             The data points (derived from the data in Ref.
             \citep{Bienayme:2005py}) and its errors are shown only for
 comparison, since these data points have not been used in the fit to the constraints mentioned in Paper I. The errors are taken to be 30\%, as estimated from the uncertainties in the phase-space distribution of the Hipparcos data and
 the parametrization of the potential.}
    \label{f3}
  \end{center}
\end{figure}
\subsection{Surface density}
The enhanced DM density from the rings leads to a different  gravitational potential in the solar neighborhood (''local'' potential), which in turn is expected to change the spatial distribution and velocity distribution of the local star populations in much the same way as the gas distribution in the earth's atmosphere is determined by its gravitational potential. So the distribution of star counts as function of galactic height above the disk  yields information on the
{\it total} local matter density, which can be compared with the visible matter density, thus yielding  information on the local DM density. The principle was pioneered by Jan Oort in the 1960's, but a seminal study using much better data was performed by Kuijken \& Gilmore (KG) \citep{Kuijken:1989,Kuijken:1989hu,Kuijken:1991mw}. They concluded that locally little dark matter was needed, a conclusion confirmed later by independent analyses using the precise data from the Hipparcos satellite \citep{Holmberg:2004fj,Bienayme:2005py}.
These data provided  also a precise measurement of the
local mass density (Oort limit), which includes visible and dark matter \citep{Holmberg:2004fj,Korchagin:2003yk}:
\begin{equation}
\rho_{\odot,tot} (z=0) = 0.102 \pm 0.005\ M_\odot\ pc^{-3}.
\label{oort2}
\end{equation}
From a comparison with Eq. \ref{oort1} one observes that the local DM density is about 30\% of the total density, which is
clearly a significant contribution. How can it be that previous authors concluded from the same star  counts  that little DM is needed?

To understand this surprising result, let us go back and see what is really measured.
The number density and velocities of a tracer population is measured as function of the distance above  the disk.
 We assume that the stars move on nearly circular orbits in the plane, so the vertical motions are independent of the circular motions. Furthermore, we assume that the  tracer population is well-mixed, i.e. the phase-space distribution $f(z,v_z)$ is in a steady-state in a static potential.  Then the energy at the local solar Galactocentric radius $R_0$ can be written as $E_z = \Phi(R_0,z) +\frac{1}{2}v_z^2$ and is an integral of motion with $E_z <0$ for bound stars.  The distribution function is given by the Boltzmann factor: $f(z,v_z)=f(E_z)\propto {\rm exp}(-E_z/\sigma(z)^2)$, where the velocity dispersion is defined by $\sigma(z)^2=<v_z^2>-<v_z>^2$. Integrating over the velocity yields for the number density of stars as function of $z$: $n(z)=n_0 ~ {\rm exp}(-\Phi(R_0,z)/\sigma(z)^2)$. So measuring the number density  $n(z)$ and the velocity dispersion $\sigma(z)$ of  stars as function of $z$ determines the local gravitational potential $\Phi(R_0,z)$ as function of $z$.

 To obtain the potential from the data it is convenient to have an analytical parametrization. KG modeled the potential as the sum of two components \citep{Kuijken:1989}:
 \begin{equation}
\Phi_z/2\pi G=\Sigma_0 \left(\sqrt{z^2+D^2}-D\right)  + \rho_{\rm  eff}z^2.
\label{oort3}
\end{equation}
 The first term is the potential of an exponential disk with a scale height D and density $\Sigma_0$, while the latter represents the potential of a locally constant halo DM density $\rho_{\rm  eff}$. If the radial and $z$ movements are independent, $\rho_{\rm  eff}$ equals the local DM density $\rho_{\odot,tot} (z=0)$, which is furthermore assumed to
 be independent of $z$ for the small volume considered.
 For precise enough data one could obtain the three parameters $\Sigma_0$, $\rho_{\rm  eff}$ and $D$ from the  distribution function.
However, these parameters are strongly correlated and the data are not  precise enough. Therefore additional measurements
are welcome. The halo density $\rho_{\rm  eff}$ can be obtained from the rotation curve, as discussed above. Furthermore, the Oort limit (Eq. \ref{oort2}) is given by the sum of
the DM density and the disk density:
 \begin{equation}
\rho_{\rm total}(z=0)=\rho_{\rm  eff}+\Sigma_0/(2\,D).
\label{oort4}
\end{equation}
This yields a relation between $\Sigma_0$ and $D$, so if $\rho_{\rm  eff}$ is obtained from the RC,
one basically has only one free parameter left, namely $D$, which can be precisely
determined from a fit to the distribution function. Unfortunately, a different choice of  $\rho_{\rm  eff}$ will yield a different
value of the scale parameter $D$ via the constraint in Eq. \ref{oort4}.
What  basically happens: $\rho(z)$ is fixed at $z=0$ by the Oort limit and for $z>>D$ by the
 $\rho_{\rm  eff}$ from the RC and the interpolation at intermediate $z$ is given by $D$.
 If we go from a RC without rings to a RC curve with rings, the values of $\rho_{\rm  eff}$ and $D$ will change. However, the  potential in Eq. \ref{oort3}, as probed by  the distribution of stars,
may change  less and be consistent with the data. To see this one can relate an axisymmetric mass distribution to the potential via the Poisson equation:
 \begin{equation}
4\pi\rho(R,z)=\nabla \Phi(R,z) = \frac{\partial^2 \Phi(R,z)}{\partial z^2}+\frac{1}{R} \frac{\partial}{\partial R}\left[R\frac{\partial \Phi(R,z)}{\partial R}\right]
\label{Poisson}
\end{equation}
The surface density is obtained by integrating the density along $z$, which is then proportional to the first derivative of the potential or the vertical force $F_z$. This force can be written in terms of the parametrization of the potential as:
\begin{equation}
  F_z/2\pi G  = \frac{\Sigma_0 z}{\sqrt{z^2+D^2}}  + 2\rho_{\rm  eff}.
\label{force}
\end{equation}
From the Poisson equation it is clear that the surface density is obtained by integrating the potential, so the surface density is proportional to the vertical force plus a small correction term, if the force varies with radius:
\begin{equation}
  \Sigma_{def}\ (< |z|) = \int\limits_{-z}^z \rho(z^\prime)\ dz^\prime= \frac{1}{2\pi G}\left(F_z-\frac{z}{R_0}\frac{\partial v_r^2}{\partial R}\right).
\label{surface}
\end{equation}
Here $v_r=R{\partial \Phi(R_0,z)}/{\partial R}$ is the local rotation speed.
If we choose $\rho_{\rm  eff}$ from the rotation curve without rings, we obtain results similar to previous measurements for $|z|=1.1$ kpc: $\Sigma_{tot}\ (< |z|)=71\pm6$  M$_\odot$~pc$^{-2}$.
However, if we choose the four times higher value of  $\rho_{\rm  eff}$  from the RC including the rings, we find
a considerably higher value:
\begin{equation}
\Sigma_{tot}\ (< |z|)=92\pm7~ M_\odot~pc^{-2}.
\label{surface1}
\end{equation}
 Thus an increase in $\rho_{\rm  eff}$
 leads to  a larger surface density, but  this changes only the {\it slope} of the potential
 for $z>>D$, not so much its value, as demonstrated in Fig. \ref{f3}. Here we simply plotted the potential as calculated
 for the mass distribution obtained from a fit to the RC with and without rings.
 For the fit the surface density  was not set to
 the value used in Paper I, but became a function of $\rho_{\rm  eff}$:
 $\Sigma_{\mathrm{tot}} = a + b \cdot \rho_{\odot,\mathrm{DM}}$
  with $a = 57.5$ M$_{\odot}$~pc$^{-2}$ and $b = 1.1 \cdot 10^3$ pc.  This relation was obtained from a fit to
  the values in Table 3 of Ref. \citep{Bienayme:2005py}.

 To check that we have a self-consistent solution the surface density was calculated  first by direct integration
 of the density using the mass model in Paper I for the visible matter and Eq. \ref{eq:HaloRings} for the DM density,
 i.e. we use the first equality in Eq. \ref{surface}.
 Secondly, the local gravitational potential was calculated from the mass distributions for visíble and DM and the vertical force was calculated from the derivative of the potential leading to the surface density using the last equality in Eq. \ref{surface}. Both methods agree within errors.

 In summary, the distribution function of local stars probe the potential, {\it not} the surface density. The latter is determined by the {\it slope} of the potential and the DM dominates only in the region  $0.5<|z| < 1.1$ kpc.
 A fourfold increase in the DM density changes the slope of the potential in this region by about 20\%, which is inside the experimental errors.
\begin{figure}
  \begin{center}
  \includegraphics[width=0.8\columnwidth,angle=0]{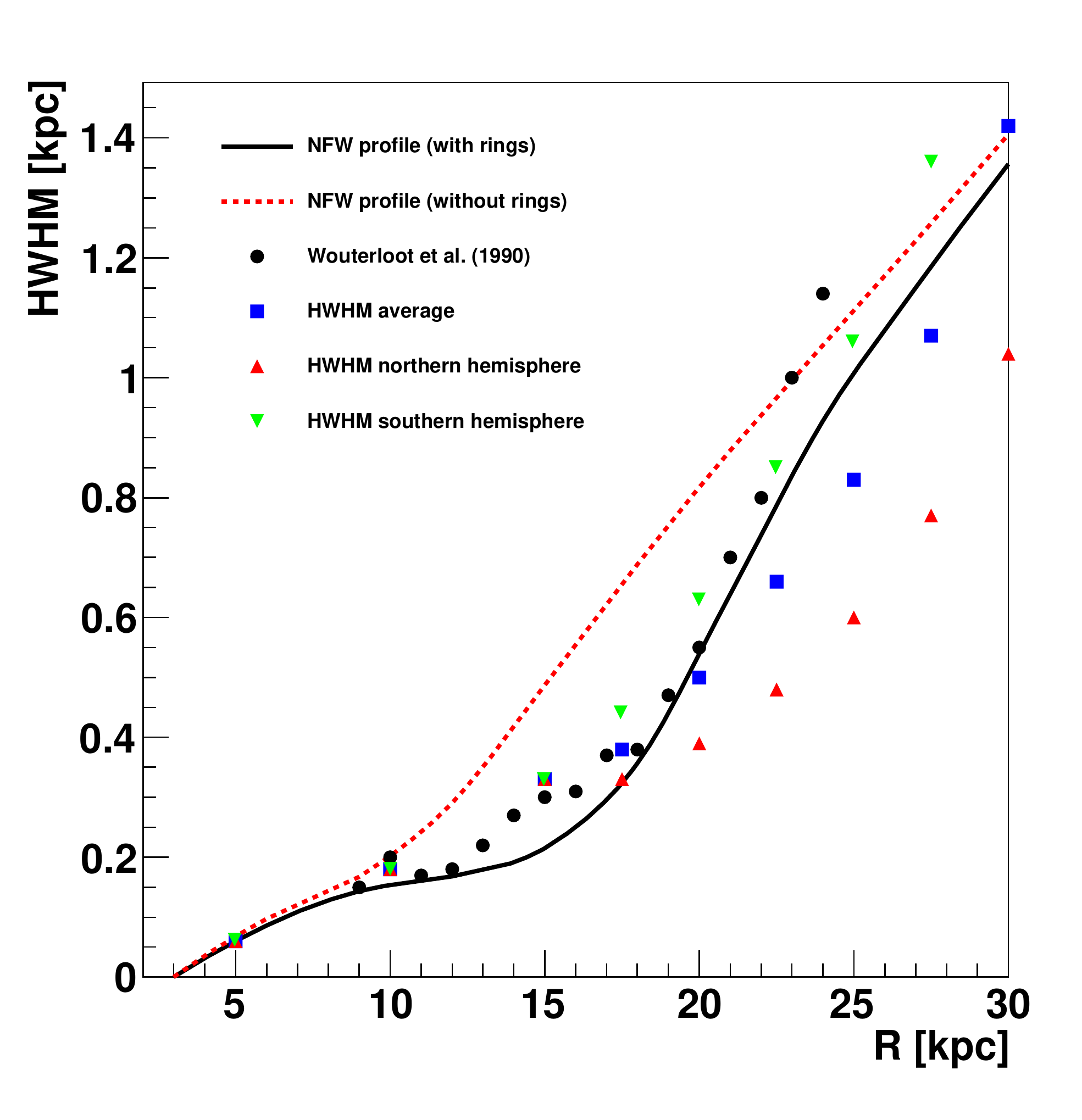}
   \caption{Gas flaring  of a DM halo profile combined
             with a substructure of two  DM rings at 4.5 and 12.5 kpc.}
    \label{f4}
  \end{center}
\end{figure}
\subsection{Gas flaring}
From the gravitational potential one can calculate the corresponding gas flaring, which
is the effect of the increase in the half-width-at-half-maximum (HWHM) of the gas layer as function of radial distance from the Galactic center due to the decrease in gravitational potential. Of course, the gas flaring
will be affected by the donut-like DM rings.
In analogy to the barometric equation the vertical decrease of the interstellar hydrogen can
be parameterized as:
\begin{equation}
  \rho(z) = \rho_{\odot,tot} \cdot \exp \left(- \frac{\Phi(z)}{\omega^2}\right).
  \label{eq:BaroEq}
\end{equation}
The experimental determination of the
velocity dispersion $\omega$ yielded values of \mbox{7 $\pm$ 1 km~s$^{-1}$}
 \citep{Shostak:1984mw} and \mbox{8 $\pm$ 1 km~s$^{-1}$} \citep{Lewis:1984mw}.
Fig. \ref{f4} shows the gas flaring for a potential with and without DM rings,
 as obtained from the measurement of the spatial distribution of the 21 cm emission line of atomic hydrogen. Note that the gas flaring is independent
 of the warping of the disk, since the center-of-gravity of the gas layer was allowed to differ from the galactic plane. The gas flaring is different for the northern
 and southern
 hemisphere \citep{Kalberla:2007sr}, which indicates that either the gravitational potential is not axisymmetric in the disk or local phenomena, like recent star bursts, cause locally additional pressure.   However, both regions are clearly  better described by the inclusion of the outer ring.
 The averaged gas flaring is better described for  an outer ring mass of $2.1 \cdot 10^{10}$~M$_\odot$, while the RC is best described by a twice heavier outer ring mass. Given the fact that the outer RC and the gas flaring use data  at very different longitudes and the northern and southern hemisphere show that the gas flaring is dependent on longitude one should not worry about a factor of two.
\section{Summary}
Up to know determinations of the dark matter density typically only considered the rotation curve of the inner Galaxy and found values around 0.3-0.4 GeV/cm$^3$.
New observations of the rotation curve challenge this result: the VERA Very Large Baseline Interferometer confirm the dip-like structure in the rotation curve around 9 kpc beyond doubt. Such a dip can only be caused by ringlike DM structures, since spherical shells do not influence the gravitational potential inside. It would be hard to explain the dip by spiral arms or warping of the disk, since  spiral arms   with such high density have not been observed in the outer Galaxy and  the VLBI measurements used maser light from molecular clouds, which is independent of the warping.

To explain the two broad ''dips'' at 3 and 9 kpc in the rotation curve  two ringlike structures are needed. A fit yields  a maximum density of the donut-like rings at 4.2 and 12.4 kpc with a Gaussian width in radius of 2.5 and 3.2 kpc, respectively. The first surprising result is not only a reasonable description of the rotation curve including the precise VLBI data, but also the fact that the position and size of these rings coincide with the dust ring at 4 kpc and the Monocerus ring of stars at 13 kpc.
The first one is suggestive of a local potential well, while the latter is thought to originate from the infall of a dwarf galaxy, as suggested by N-body simulations.  Of course, it is worth emphasizing that  our model of circular ringlike structures should not be taken literally, since DM structures will not be perfectly circular. However, the data are most affected by the nearby segment of the structure, which seems well described by a circular part, at least changing the ellipticity of the rings did not modify the results  and both, the Monocerus ring and dust ring are largely circular as well. It should  be noticed that the significant enhancement of the rings in comparison with  the smooth halo in Fig. \ref{f2} may be exaggerated, since recent N-body simulations  predict an enhancement of the smooth halo in the disk, the so-called dark disks \citep{Purcell:2009yp}, which would reduce the contrast. Furthermore, the parametrization with ringlike structures is certainly an oversimplification, so the error on the local DM density is large and the difference between DM density with and without rings is  about three standard deviations.

The second surprising result is the agreement with the s-shaped gas flaring in the outer Galaxy, which also needs a
ringlike structure with a similar mass at a similar position in the outer Galaxy. The ''dip'' in the gas flaring was determined in such a way
that it is independent of the warping, basically because the center  of the gas layer is not required to be at z=0.

 The third surprising result is that  a fourfold increase of the local DM density  compared to a fit without rings is still in excellent agreement with the local surface density, as determined from the density distribution and velocity dispersion of the local stars. From these measurements it was claimed that our Galaxy needs locally no DM in agreement with the 1/$\sqrt{r}$ decrease of the inner rotation curve, which is suggestive of a potential determined by visible matter alone.   If the decreasing rotation curve is attributed to the negative contribution from the outer ring, one obtains a larger local DM density and a correspondingly higher surface density. A higher surface density implies a higher vertical force, which only changes the slope of the gravitational potential, not its magnitude. However, its magnitude is what is probed by the distribution function of local stars. Below $z$=0.5 kpc  the potential is dominated by visible matter, so    a fourfold increase in the DM density will change the slope of the potential smoothly by about 20\%, but the data at large $z$ are not precise enough to be sensitive to such a moderate change of slope.

  In summary, the higher DM density found in this paper seems inevitably required by the structure of the rotation curve using the new VLBI measurements {\it and } the gas flaring and was found to be perfectly compatible with the surface density.  Such a possible increase of the DM density in the disk is of prime importance for  DM searches, both direct and indirect, since  the direct detection rate is  proportional to the local DM density, while  nearby ringlike DM structures would  change the directional dependence for gamma rays in indirect DM searches.

\section{Acknowledgments}
Support from the Deutsche Luft und Raumfahrt (DLR) and the Bundesministerium for Bildung und Forschung (BMBF) is gratefully acknowledged. We thank the anonymous referee for helpful suggestions to improve the manuscript regarding the surface density.
%
%

\begin{thebibliography}{41}
\providecommand{\natexlab}[1]{#1}
\providecommand{\url}[1]{\texttt{#1}}
\expandafter\ifx\csname urlstyle\endcsname\relax
  \providecommand{\doi}[1]{doi: #1}\else
  \providecommand{\doi}{doi: \begingroup \urlstyle{rm}\Url}\fi

\bibitem[Fich and Tremaine(1991)]{Fich:1991ej}
M.~Fich and S.~Tremaine,
\newblock {The mass of the galaxy},
\newblock \emph{Ann. Rev. Astron. Astrophys.}, 29:\penalty0 409--445, 1991.

\bibitem[Sofue and Rubin(2001)]{Sofue:2000jx}
Y.~Sofue and V.~Rubin,
\newblock {Rotation Curves of Spiral Galaxies},
\newblock \emph{Ann. Rev. Astron. Astrophys.}, 39:\penalty0 137--174, 2001.

\bibitem[Sofue et~al.(2009)Sofue, Honma, and Omodaka]{Sofue:2008wt}
Y.~Sofue, M.~Honma, and T.~Omodaka,
\newblock {Unified Rotation Curve of the Galaxy -- Decomposition into de
  Vaucouleurs Bulge, Disk, Dark Halo, and the 9-kpc Rotation Dip --},
\newblock \emph{Publ. Astron. Soc. Jap.}, 61:\penalty0 229, 2009.

\bibitem[Binney and Dehnen(1997)]{Binney:1996fb}
J.~Binney and W.~Dehnen,
\newblock {The outer rotation curve of the Milky Way},
\newblock \emph{Mon. Not. Roy. Astron. Soc.}, 287:\penalty0 L5, 1997.

\bibitem[Catena and Ullio(2010)]{Catena:2009mf}
R.~Catena and P.~Ullio,
\newblock {A novel determination of the local dark matter density},
\newblock \emph{JCAP}, 1008:\penalty0 004, 2010.

\bibitem[Weber and de~Boer(2010)]{Weber:2009pt}
M.~Weber and W.~de~Boer,
\newblock {Determination of the Local Dark Matter Density in our Galaxy},
\newblock \emph{Astron. Astrophys.}, 509:\penalty0 A25, 2010.

\bibitem[Salucci et~al.(2010)Salucci, Nesti, Gentile, and
  Martins]{Salucci:2010qr}
P.~Salucci, F.~Nesti, G.~Gentile, and C.~F. Martins,
\newblock {The dark matter density at the Sun's location},
\newblock \emph{Astron. Astrophys.}, 523:\penalty0 A83, 2010.

\bibitem[Korchagin et~al.(2003)Korchagin, Girard, Borkova, Dinescu, and van
  Altena]{Korchagin:2003yk}
V.~I. Korchagin, Terrence~M. Girard, T.~V. Borkova, D.~I. Dinescu, and W.~F.
  van Altena,
\newblock {Local Surface Density of the Galactic Disk from a 3-D Stellar
  Velocity Sample},
\newblock 2003.

\bibitem[Holmberg and Flynn(2004)]{Holmberg:2004fj}
J.~Holmberg and C.~Flynn,
\newblock {The local surface density of disc matter mapped by Hipparcos},
\newblock \emph{Mon. Not. Roy. Astron. Soc.}, 352:\penalty0 440, 2004.

\bibitem[Bienayme et~al.(2006)Bienayme, Soubiran, Mishenina, Kovtyukh, and
  Siebert]{Bienayme:2005py}
O.~Bienayme, C.~Soubiran, T.~V. Mishenina, V.~V. Kovtyukh, and A.~Siebert,
\newblock {Vertical distribution of Galactic disk stars: III. The Galactic disk
  surface mass density from red clump giants},
\newblock \emph{Astron. Astrophys.}, 446:\penalty0 933--942, 2006.

\bibitem[Kuijken and Gilmore(1989{\natexlab{a}})]{Kuijken:1989}
K.~Kuijken and G.~Gilmore,
\newblock {The Mass Distribution in the Galactic Disc - Part One - A Technique
  to determine the Integral Surface Mass Density of the Disc near the Sun},
\newblock \emph{Mon. Not. Roy. Astron. Soc.}, 239:\penalty0 571,
  1989{\natexlab{a}}.

\bibitem[Kuijken and Gilmore(1989{\natexlab{b}})]{Kuijken:1989hu}
K.~Kuijken and G.~Gilmore,
\newblock {The Mass Distribution in the Galactic Disc - Part Two -
  Determination of the Surface Mass Density of the Galactic Disc Near the Sun},
\newblock \emph{Mon. Not. Roy. Astron. Soc.}, 239:\penalty0 605,
  1989{\natexlab{b}}.

\bibitem[Kuijken and Gilmore(1991)]{Kuijken:1991mw}
K.~Kuijken and G.~Gilmore,
\newblock {The galactic disk surface mass density and the Galactic force K(z)
  at Z = 1.1 kiloparsecs},
\newblock \emph{Astrophys. J.}, 367:\penalty0 L9--L13, 1991.

\bibitem[Oh et~al.(2010)]{Oh:2010zz}
C.~S. Oh et~al.,
\newblock {{VERA} Observations of H$_2$O Maser Sources in Three Massive
  Star-Forming Regions and Galactic Rotation Measurements},
\newblock \emph{Publ. Astron. Soc. Jap.}, 62:\penalty0 101--114, 2010.

\bibitem[Kalberla et~al.(2007)Kalberla, Dedes, Kerp, and Haud]{Kalberla:2007sr}
P.~M.~W. Kalberla, L.~Dedes, J.~Kerp, and U.~Haud,
\newblock {Dark matter in the Milky Way, II. the HI gas distribution as a
  tracer of the gravitational potential},
\newblock \emph{Astron. Astrophys.}, 469:\penalty0 511--527, 2007.

\bibitem[Newberg et~al.(2002)]{Newberg:2001sx}
H.~J. Newberg et~al.,
\newblock {The Ghost of Sagittarius and Lumps in the Halo of the Milky Way},
\newblock \emph{Astrophys. J.}, 569:\penalty0 245--274, 2002.

\bibitem[Yanny et~al.(2003)]{Yanny:2003zu}
B.~Yanny et~al.,
\newblock {A Low Latitude Halo Stream around the Milky Way},
\newblock \emph{Astrophys. J.}, 588:\penalty0 824, 2003.

\bibitem[Ibata et~al.(2003)Ibata, Irwin, Lewis, Ferguson, and
  Tanvir]{Ibata:2003di}
R.~A. Ibata, M.~J. Irwin, G.~F. Lewis, A.~M.~N. Ferguson, and N.~Tanvir,
\newblock {One Ring to Encompass them All: A giant stellar structure that
  surrounds the Galaxy},
\newblock \emph{Mon. Not. Roy. Astron. Soc.}, 340:\penalty0 L21, 2003.

\bibitem[Rocha-Pinto et~al.(2004)Rocha-Pinto, Majewski, Skrutskie, Crane, and
  Patterson]{RochaPinto:2004ru}
H.~J. Rocha-Pinto, S.~R. Majewski, M.~F. Skrutskie, Je.~D. Crane, and R.~J.
  Patterson,
\newblock {Exploring Halo Substructure with Giant Stars: A diffuse star cloud
  or tidal debris around the Milky Way in Triangulum-Andromeda},
\newblock \emph{Astrophys. J.}, 615:\penalty0 732--737, 2004.

\bibitem[Martin et~al.(2005)]{Martin:2005xa}
N.~F. Martin et~al.,
\newblock {A radial velocity survey of low Galactic latitude structures: I.
  Kinematics of the Canis Major dwarf galaxy},
\newblock \emph{Mon. Not. Roy. Astron. Soc.}, 362:\penalty0 906--914, 2005.

\bibitem[Conn et~al.(2005{\natexlab{a}})]{Conn:2005xd}
B.~C. Conn et~al.,
\newblock {A radial velocity survey of low Galactic latitude structures: II.
  The Monoceros Ring behind the Canis Major dwarf galaxy},
\newblock \emph{Mon. Not. Roy. Astron. Soc. Lett.}, 364:\penalty0 L13--L17,
  2005{\natexlab{a}}.

\bibitem[Martin et~al.(2006)]{Martin:2006rg}
N.~F. Martin et~al.,
\newblock {A radial velocity survey of low Galactic latitude structures: III.
  The Monoceros Ring in front of the Carina and Andromeda galaxies},
\newblock \emph{Mon. Not. Roy. Astron. Soc. Lett.}, 367:\penalty0 L69--L73,
  2006.

\bibitem[Martin et~al.(2004)]{Martin:2003vk}
N.~F. Martin et~al.,
\newblock {A dwarf galaxy remnant in Canis Major: the fossil of an in-plane
  accretion onto the Milky Way},
\newblock \emph{Mon. Not. Roy. Astron. Soc.}, 348:\penalty0 12, 2004.

\bibitem[Penarrubia et~al.(2005)]{Penarrubia:2004qw}
J.~Penarrubia et~al.,
\newblock {A comprehensive model for the Monoceros tidal stream},
\newblock \emph{Astrophys. J.}, 626:\penalty0 128--144, 2005.

\bibitem[Kazantzidis et~al.(2008)Kazantzidis, Bullock, Zentner, Kravtsov, and
  Moustakas]{Kazantzidis:2007hy}
S.~Kazantzidis, J.~S. Bullock, A.~R. Zentner, A.~V. Kravtsov, and L.~A.
  Moustakas,
\newblock {Cold Dark Matter Substructure and Galactic Disks I: Morphological
  Signatures of Hierarchical Satellite Accretion},
\newblock \emph{Astrophys. J.}, 688:\penalty0 254--276, 2008.

\bibitem[Conn et~al.(2005{\natexlab{b}})]{Conn:2005cn}
B.~C. Conn et~al.,
\newblock {The INT/WFC survey of the Monoceros Ring: Accretion origin or
  Galactic Anomaly?}
\newblock \emph{Mon. Not. Roy. Astron. Soc.}, 362:\penalty0 475--488,
  2005{\natexlab{b}}.

\bibitem[Moitinho et~al.(2006)]{Moitinho:2006ru}
A.~Moitinho et~al.,
\newblock {Spiral structure of the Third Galactic Quadrant and the solution to
  the Canis Major debate},
\newblock \emph{Mon. Not. Roy. Astron. Soc. Lett.}, 368:\penalty0 L77, 2006.

\bibitem[Momany et~al.(2006)]{Momany:2006ch}
Y.~Momany et~al.,
\newblock {Outer structure of the Galactic warp and flare: explaining the Canis
  Major over-density},
\newblock \emph{Astron. Astrophys.}, 451:\penalty0 515--538, 2006.

\bibitem[Wyse et~al.(2006)]{Wyse:2006xz}
R.~F.~G. Wyse et~al.,
\newblock {Further Evidence for a Merger Origin for the Thick Disk: Galactic
  Stars Along Lines-of-sight to Dwarf Spheroidal Galaxies},
\newblock \emph{Astrophys. J.}, 639:\penalty0 L13--L16, 2006.

\bibitem[Ruchti et~al.(2010)]{Ruchti:2010cz}
G.~R. Ruchti et~al,
\newblock {Origins of the Thick Disk as Traced by the Alpha-Elements of
  Metal-Poor Giant Stars Selected from RAVE},
\newblock \emph{Astrophysical Journal}, Lett. 721:\penalty0 L92--L96, 2010.

\bibitem[Robin et~al.(1996)Robin, Haywood, Creze, Ojha, and
  Bienayme]{Robin:1995fr}
A.~C. Robin, M.~Haywood, M.~Creze, D.~K. Ojha, and O.~Bienayme.
\newblock {The thick disc of the galaxy: Sequel of a merging event}.
\newblock \emph{Astron. Astrophys.}, 305:\penalty0 125, 1996.

\bibitem[Dierickx et~al.(2010)Dierickx, Klement, Rix, and Liu]{Dierickx:2010jc}
M.~Dierickx, Rainer~J. Klement, H.-W. Rix, and Chao Liu,
\newblock {Observational Evidence from SDSS for a Merger Origin of the Milky
  Way's Thick Disk},
\newblock arXiv 1009.1616, 2010.

\bibitem[Kazantzidis et~al.(2009)Kazantzidis, Zentner, Kravtsov, Bullock, and
  Debattista]{Kazantzidis:2009zq}
S.~Kazantzidis, A.~R. Zentner, A.~V. Kravtsov, J.~S. Bullock, and V.~P.
  Debattista,
\newblock {Cold Dark Matter Substructure and Galactic Disks II: Dynamical
  Effects of Hierarchical Satellite Accretion},
\newblock \emph{Astrophys. J.}, 700:\penalty0 1896--1920, 2009.

\bibitem[Read et~al.(2009)Read, Mayer, Brooks, Governato, and
  Lake]{Read:2009iv}
J.~I. Read, L.~Mayer, A.~M. Brooks, F.~Governato, and G.~Lake,
\newblock {A dark matter disc in three cosmological simulations of Milky Way
  mass galaxies},
\newblock \emph{Mon. Not. Roy. Astron. Soc.}, 397:\penalty0 44, 2009.

\bibitem[Drimmel and Spergel(2001)]{Drimmel:2001ti}
R.~Drimmel and D.~N. Spergel,
\newblock {Three Dimensional Structure of the Milky Way Disk},
\newblock \emph{Astrophys. J.}, 556:\penalty0 181, 2001.

\bibitem[Wang et~al.(2010)]{Wang:2010ik}
J.~Wang et~al,
\newblock {Assembly History and Structure of Galactic Cold Dark Matter Halos},
\newblock to be published, arXiv 1008.5114, 2010.

\bibitem[Villalobos et~al.(2010)Villalobos, Kazantzidis, and
  Helmi]{Villalobos:2009qc}
A.~Villalobos, S.~Kazantzidis, and A.~Helmi,
\newblock {Thick-disk evolution induced by the growth of an embedded thin
  disk},
\newblock \emph{Astrophys. J.}, 718:\penalty0 314--330, 2010.

\bibitem[Ibata et~al.(2005)]{Ibata:2005bd}
R.~Ibata et~al.,
\newblock {On the accretion origin of a vast extended stellar disk around the
  Andromeda galaxy},
\newblock \emph{Astrophys. J.}, 634:\penalty0 287--313, 2005.

\bibitem[Shostak and van~der Kruit(1984)]{Shostak:1984mw}
G.~S. Shostak and P.~C. van~der Kruit,
\newblock {Studies of nearly face-on spiral galaxies. II - H I synthesis observations and optical surface photometry of NGC 628}
\newblock \emph{Astron. Astrophys.}, 132:\penalty0 20--32, 1984.

\bibitem[Lewis(1984)]{Lewis:1984mw}
B.~M. Lewis,
\newblock {Face-on galaxies},
\newblock \emph{Astrophys. J.}, 285:\penalty0 453--457, 1984.

\bibitem[Purcell et~al.(2009)Purcell, Bullock, and Kaplinghat]{Purcell:2009yp}
C.~W. Purcell, J.~S. Bullock, and M.~Kaplinghat,
\newblock {The Dark Disk of the Milky Way},
\newblock \emph{Astrophys. J.}, 703:\penalty0 2275--2284, 2009.

\end{thebibliography}

\end{document}